\documentclass[12pt,preprint]{aastex}

\usepackage{amsmath}
\usepackage{array}
\usepackage{wasysym}
\usepackage{graphicx, subfigure}

\shorttitle{Presolar grains from post-AGB stars}
\shortauthors{Jadhav et al.}

\begin{document}

\title{Relics of ancient post-AGB stars in a primitive meteorite}

\author{M. Jadhav\altaffilmark{1}}
\affil{$^{1}$Hawai\textquoteleft i Institute of Geophysics and Planetology, University of Hawai\textquoteleft i at M\={a}noa, Honolulu, HI 96822, USA}
\email{manavi@higp.hawaii.edu}

\author{M. Pignatari\altaffilmark{2,\dagger}}
\affil{$^{2}$Department of Physics, University of Basel, CH-4056 Basel, Switzerland}

\author{F. Herwig\altaffilmark{3,\dagger,\ddagger}}
\affil{$^{3}$Department of Physics \& Astronomy, University of Victoria, Victoria, BC V8P5C2, Canada}

\author{E. Zinner\altaffilmark{4}}
\affil{$^{4}$Laboratory for Space Sciences \& the Physics Department, Washington University, St. Louis, MO 63130, USA}

\author{R. Gallino\altaffilmark{5}}
\affil{$^{5}$Diparttimento di Fisica Generale, Università di Torino \& INAF, Osservatorio di Teramo}

\and

\author{G. R. Huss\altaffilmark{1}}
\affil{$^{1}$Hawai\textquoteleft i Institute of Geophysics and Planetology, University of Hawai\textquoteleft i at M\={a}noa, Honolulu, HI 96822}

Accepted for publication in Ap.J. Letters, October 4, 2013
\altaffiltext{$\dagger$}{NuGrid Collaboration: http://www.nugridstars.org}
\altaffiltext{$\ddagger$}{The Joint Institute for Nuclear Astrophysics, Notre Dame, IN 46556, USA}

\begin{abstract}
Graphite is one of the many presolar circumstellar condensate species found in primitive meteorites. While the isotopic compositions of low-density graphite grains indicate an origin in core-collapse supernovae, some high-density grains have extreme isotopic anomalies in C, Ca and Ti, which cannot be explained by envelope predictions of asymptotic giant branch (AGB) stars or theoretical supernova models. The Ca and Ti isotopic anomalies, however, match the predictions of He-shell abundances in AGB stars. In this study, we show that the C, Ca, and Ti isotopic anomalies are consistent with nucleosynthesis predictions of the H-ingestion phase during a very late thermal pulse (VLTP) event in post-AGB stars. The low $^{12}$C/$^{13}$C isotopic ratios in these grains are a result of abundant $^{12}$C efficiently capturing the protons that are being ingested during the VLTP. Very high neutron densities of $\sim 10^{15}$ cm$^{-3}$, typical of the $i$-process, are achieved during this phase in post-AGB stars. The large $^{42,43,44}$Ca excesses in some graphite grains are indicative of neutron capture nucleosynthesis during VLTP. The comparison of VLTP nucleosynthesis calculations to the graphite data also indicate that apparent anomalies in the Ti isotopic ratios are due to large contributions from $^{46,48}$Ca, which cannot be resolved from the isobars $^{46,48}$Ti during the measurements. We conclude that presolar graphite grains with moderate to extreme Ca and Ti isotopic anomalies originate in post-AGB stars that suffer a very late thermal pulse.
\end{abstract}


\section{Introduction}
\begin{figure}[h]
       \centering
        \includegraphics[width=30pc]{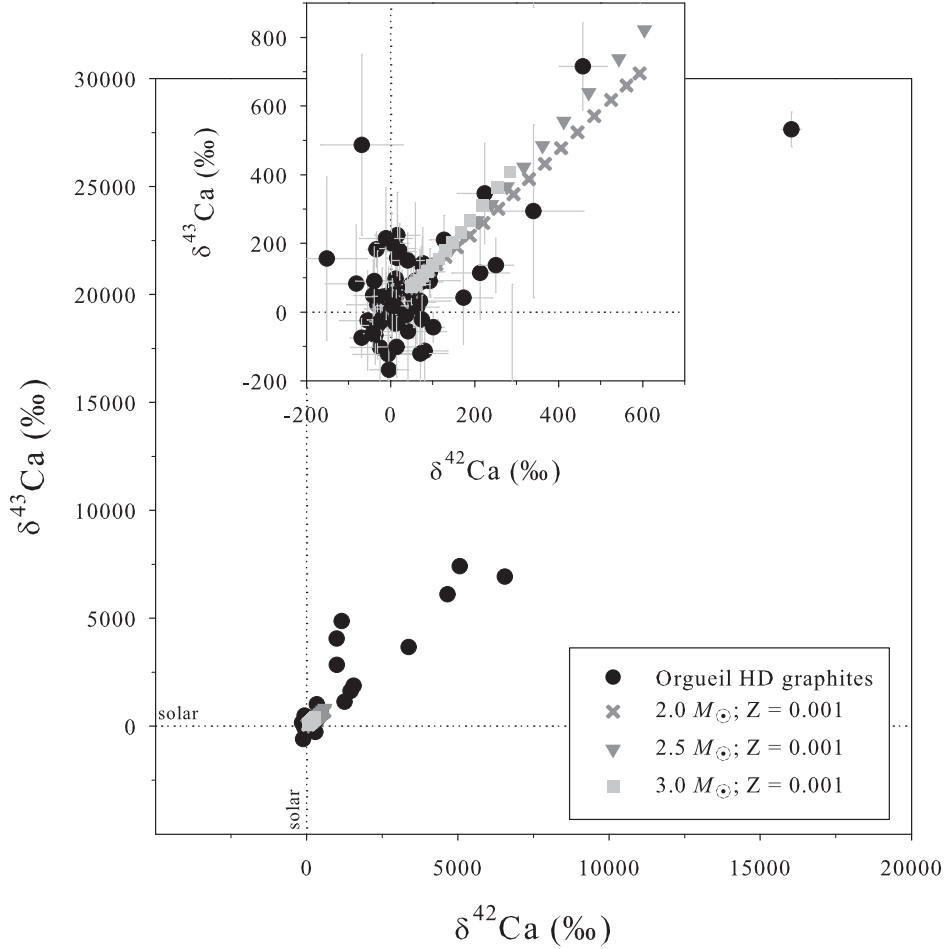}
				\figurenum{1}
        \caption{Three-isotope plot of $^{42,43}$Ca/$^{40}$Ca isotopic ratios measured in HD graphite grains. The ratios are plotted as $\delta$-values, deviations from the terrestrial ratios in permil (\permil). A majority of HD graphite grains are known to have originated in sub-solar metallicity AGB stars (e.g., Jadhav et al. 2013). Hence, the isotopic ratios in the grains are compared to model predictions for envelopes of low-metallicity AGB stars from the F.R.U.I.T.Y. database \citep{cristallo2011}. Higher metallicity models yield even smaller Ca anomalies than the $Z=0.001$ models shown here. AGB models cannot explain the extremely large Ca isotopic anomalies measured in HD graphite grains. 
				Error bars are 1$\sigma$. Dashed lines indicate solar ratios.
				\label{ca-fruity}}
 \end{figure}

Presolar graphite grains found in meteorites are now known to have several stellar sources. A majority of low-density (LD) graphite grains from the primitive meteorites Murchison and Orgueil exhibit isotopic evidence for an origin in core-collapse supernovae (CCSNe), similar to SiC-X grains \citep{amari1992}. They have large, and often correlated, excesses in $^{15}$N, $^{18}$O, and $^{28}$Si \citep{amari1995a,jadhav2006,travaglio1999}, and high inferred $^{26}$Al/$^{27}$Al, $^{41}$Ca/$^{40}$Ca, and $^{44}$Ti/$^{48}$Ti ratios \citep{nittler1996,amari1996,jadhav2013}. High-density (HD) graphite grains, on the other hand, have several stellar sources. One of the s-process components of Kr, Kr-SH, with a high $^{86}$Kr/$^{82}$Kr ratio, resides in the HD fraction of Murchison graphites and is believed to originate from AGB stars \citep{amari1995b,amari2012}. A majority of HD graphite grains from Murchison and Orgueil have correlated high $^{12}$C/$^{13}$C ratios and $^{30}$Si excesses. Such isotopic signatures are predicted for subsolar-metallicity asymptotic giant branch (AGB) stars where more $^{12}$C and $^{29,30}$Si from the He shell is dredged up into the envelope during the thermally pulsing phase than in stars of solar metallicity \citep{zinner2006}. Molecular spectral features of circumstellar dust around AGB stars indicate that the abundance of SiC grains in circumstellar dust shells decreases with decreasing stellar metallicity and is explained by the increasing C/O ratio \citep{leisenring2008}, which favors graphite condensation over SiC. Model calculations of dust formation by \citet{gail2009} also predict that AGB stars with low metallicities ($\sim$ 0.3$Z_{\odot}$) mostly produce graphite grains. Although HD graphites do not contain $^{15}$N, $^{18}$O, and $^{28}$Si excesses, \citet{jadhav2008} and \citet{jadhav2013} found HD presolar graphite grains with evidence for the initial presence of the short-lived radionuclide $^{44}$Ti, indicating an origin in CCSNe. In addition, \citet{jadhav2008} found extremely anomalous Ca and Ti isotopic ratios in some grains with $^{12}$C/$^{13}$C ratios $<$ 20. Figure \ref{ca-fruity} compares Ca isotopic data measured in HD graphites to nucleosynthesis predictions for the envelopes of low-metallicity AGB stars from the FRANEC Repository of Updated Tables and Yields (F.R.U.I.T.Y.) database \citep{cristallo2011}. The anomalies in many grains are much too large to be explained by the F.R.U.I.T.Y. envelope predictions. The Ca and Ti anomalies in these graphites match predicted, pure He-shell abundances in low-mass AGB stars. However, the very low $^{12}$C/$^{13}$C ratios of the grains with large Ca anomalies, indicative of H-burning, disagree with the almost pure $^{12}$C predicted for the He intershell, and with predictions of envelope abundances from baseline AGB models. \citet{jadhav2008} concluded that born-again AGB stars or post-AGB stars that have suffered a very late thermal pulse (VLTP) are likely sources of such graphite grains. Born-again AGB stars have previously been suggested as stellar sources for SiC-AB grains, which have $^{12}$C/$^{13}$C $<$ 10 \citep{amari2001}.

In order to test the hypothesis that $^{13}$C-enriched graphite grains with extreme Ca and Ti anomalies originate from post-AGB stars, we present here a comparison of grain data with VLTP nucleosynthesis model calculations from \citet{herwig2011}.

\section{Grain data and nucleosynthesis calculations}
\subsection{High-density, graphite grains from Orgueil}
We use C, N, Ca, and Ti isotopic data for presolar graphite grains ($>$2 $\mu$m) previously obtained by \citet{jadhav2008} (OR1f2m, 44 grains) and \citet{jadhav2011} (OR1f3m, 39 grains) from the HD fraction of Orgueil, OR1f (2.02 -- 2.04 g cm$^{-3}$). Experimental details of the measurements and isotopic characteristics of the grains are discussed in the respective references. About 8 -- 9 \% of the grains have $^{12}$C/$^{13}$C $<$ 20 and some of these grains have extreme Ca and Ti isotopic anomalies (deviations from solar ratios) that cannot be explained by nucleosynthesis calculations for envelopes of AGB stars or C-rich SN ejecta. Some grains have $^{12}$C/$^{13}$C $>$ solar (89) and moderate Ca and Ti anomalies that also exceed the isotopic ratios obtained by theoretical AGB or SN models.  
\subsection{H-ingestion post-AGB models by \citet{herwig2011}}
During their descent along the white dwarf cooling track and after H-burning has ceased, post-AGB stars that undergo a very late He flash (e.g., Sakurai's object (V4334Sgr), V 605 Aql (Nova Aql 1919), PG 1159 stars) may simultaneously exhibit signatures of neutron capture processes on several elements and low $^{12}$C/$^{13}$C ratios. At this stage in its evolution, the remnant star has lost most of its envelope, has passed through the planetary nebula phase, and has only a thin residual H layer left ($M \sim 10^{-3}-10^{-4} M_{\odot}$, e.g., Herwig et al. 2011). During the VLTP, the He intershell ignites and becomes convective. The residual H on the surface is ingested into the convection zone powered by the He-flash, which results in H-burning in hot $^{12}$C-rich layers. This reduces the $^{12}$C/$^{13}$C ratio as $^{12}$C is converted to $^{13}$C via $^{12}$C($p$,$\gamma$)$^{13}$N($\beta^{+}$)$^{13}$C. The thin, residual envelope has a limited supply of protons, which prevents the CN cycle from reaching equilibrium. The H is exhausted before $^{14}$N becomes more abundant than $^{12}$C, and the envelope remains C-rich \citep{herwig1999,herwig2001a}. With time, as H is mixed into hotter regions of the He-shell flash convection zone and the reaction rate of $^{12}$C($p$,$\gamma$)$^{13}$N increases, a thin radiative zone forms that splits the top H-burning convection zone from the He-shell burning zone below. This separation of the He intershell region is not instantaneous and allows minimal mixing of material between the zones \citep{herwig2011}. After H-ingestion begins, it takes a few hours to a day for this split to occur. At the same time, the observed heavy element abundances require that $^{13}$C burns via the $^{13}$C($\alpha,n$)$^{16}$O reaction deep at the bottom of the He intershell, resulting in neutron densities that reach $\sim$ 10$^{15}$ cm$^{-3}$, typical of the $i$-process \citep{cowan1977}. Under these conditions, the isotopic and elemental abundances in the He intershell from the previous AGB phase are modified by proton captures and the $i$-process, eventually becoming distinctly different from the He intershell abundances of AGB stars \citep{herwig2011}.

We compare C, N, Ca, and Ti isotopic data for Orgueil HD (OR1f) presolar graphite grains \citep{jadhav2008,jadhav2011} with VLTP model predictions by \citet{herwig2011}. Their calculations were guided by the elemental abundances observed in the born-again AGB star Sakurai's object (V4334 Sagittarii) \citep{asplund1999,duerbeck2000}. While Sakurai's object is not the stellar source of the HD graphites discussed in this paper, it is safe to argue that proton capture and $i$-process signatures are typical nucleosynthesis features of all VLTP events. We compare grain data with two simulations (RUN48 and RUN106) carried out for a progenitor AGB star of initial mass, $M=2 M_{\odot}$ and metallicity $Z=0.01$. In RUN48 the amount of H ingested is $X$(H) $= 5 \times 10^{-4}$ with the imposed split in the convection zone at 0.5885 $M_{\odot}$ and at $t=950$ minutes. After the split occurs, no mixing was allowed between the H- and He-burning regions. In model RUN106, the split was imposed at 1200 minutes. 

\section{Comparison of grain data and results of model calculations}
\begin{figure}[h]
       \centering
        \includegraphics[width=\textwidth]{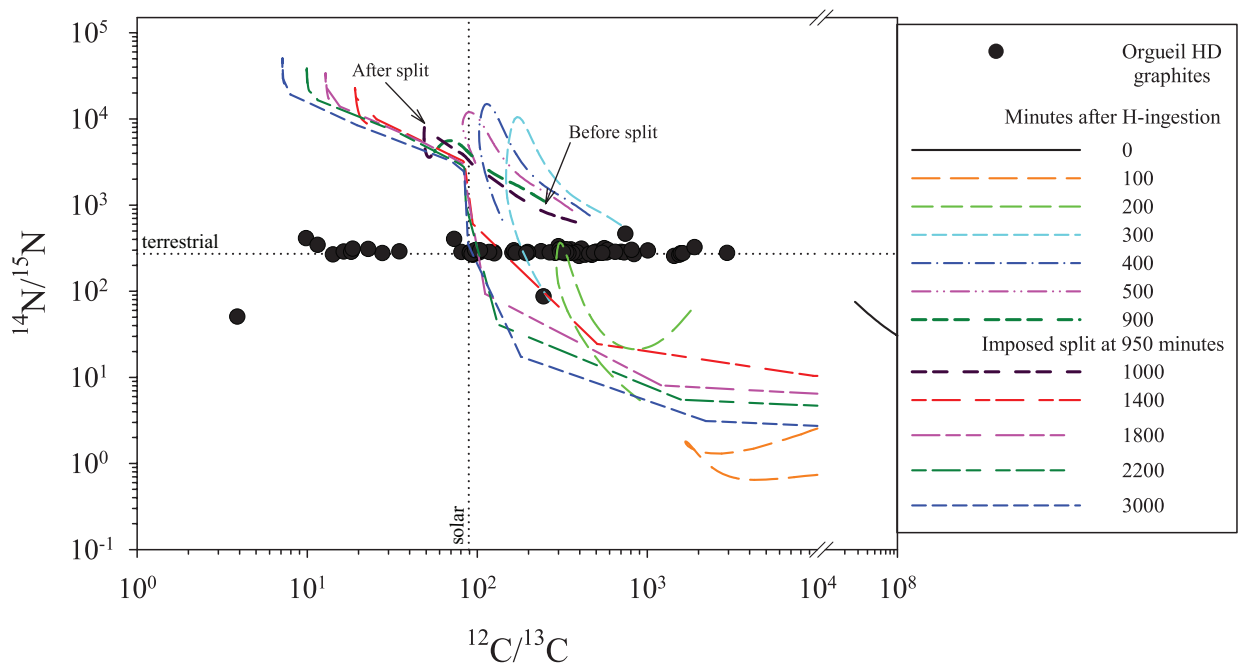}
				\figurenum{2}
       \caption{Carbon and nitrogen isotopic ratios of HD graphite grains compared to VLTP model predictions by \citet{herwig2011}. Each line, in this and all the following figures, represents an isotopic depth profile calculated by the \citet{herwig2011} VLTP models of the He intershell at a given time after the start of H-ingestion. In this simulation (RUN48), the split was imposed at 950 minutes after H-ingestion. Both H- and He-burning signatures in C and N are obtained during VLTP nucleosynthesis and predicted ratios span the entire range of C isotopic ratios obtained in HD graphite grains. The N isotopic compositions in HD graphites are believed to be equilibrated with terrestrial N and are not indicative of the stellar source of the grains.
			Error bars are 1$\sigma$. Dashed lines indicate solar (C) and terrestrial (N) ratios.
			\label{cn-vltp}}
 \end{figure}

High-density graphite grain data are compared to the VLTP models described above in Figures \ref{cn-vltp}--5. Each line in these figures represents an isotopic depth profile of the He intershell at a given time after the start of H-ingestion. Figure \ref{cn-vltp} shows the temporal evolution of C and N isotopic ratios in the intershell region for RUN48 before and after the time the He intershell region splits at 950 minutes, up to 3000 minutes, when H-ingestion is complete. The figure clearly shows the splitting of the He intershell region into H- and He-burning zones. After the split is formed, the upper H-burning zone has low $^{12}$C/$^{13}$C and high $^{14}$N/$^{15}$N ratios, while the deeper He-flash driven zone has high $^{12}$C/$^{13}$C and low $^{14}$N/$^{15}$N ratios. Particularly, this region now cannot accumulate $^{13}$C and $^{14}$N because they are rapidly depleted by alpha-capture due to the higher temperatures. Thus, the theoretical $^{12}$C/$^{13}$C ratios span the entire range observed in all HD graphites. Unfortunately, HD graphites have close-to-terrestrial $^{14}$N/$^{15}$N ratios that indicate equilibration with terrestrial N and are, therefore, not indicative of the $^{14}$N/$^{15}$N ratios that the grains inherited from their stellar source(s) \citep{jadhav2006,jadhav2013}. The high neutron densities reached during H-ingestion due to the activation of the $^{13}$C($\alpha,n$)$^{16}$O reaction, result in the efficient production of neutron-rich species as shown by \citet{herwig2011}.

\begin{figure}[h]
       \centering
        \includegraphics[width=35pc]{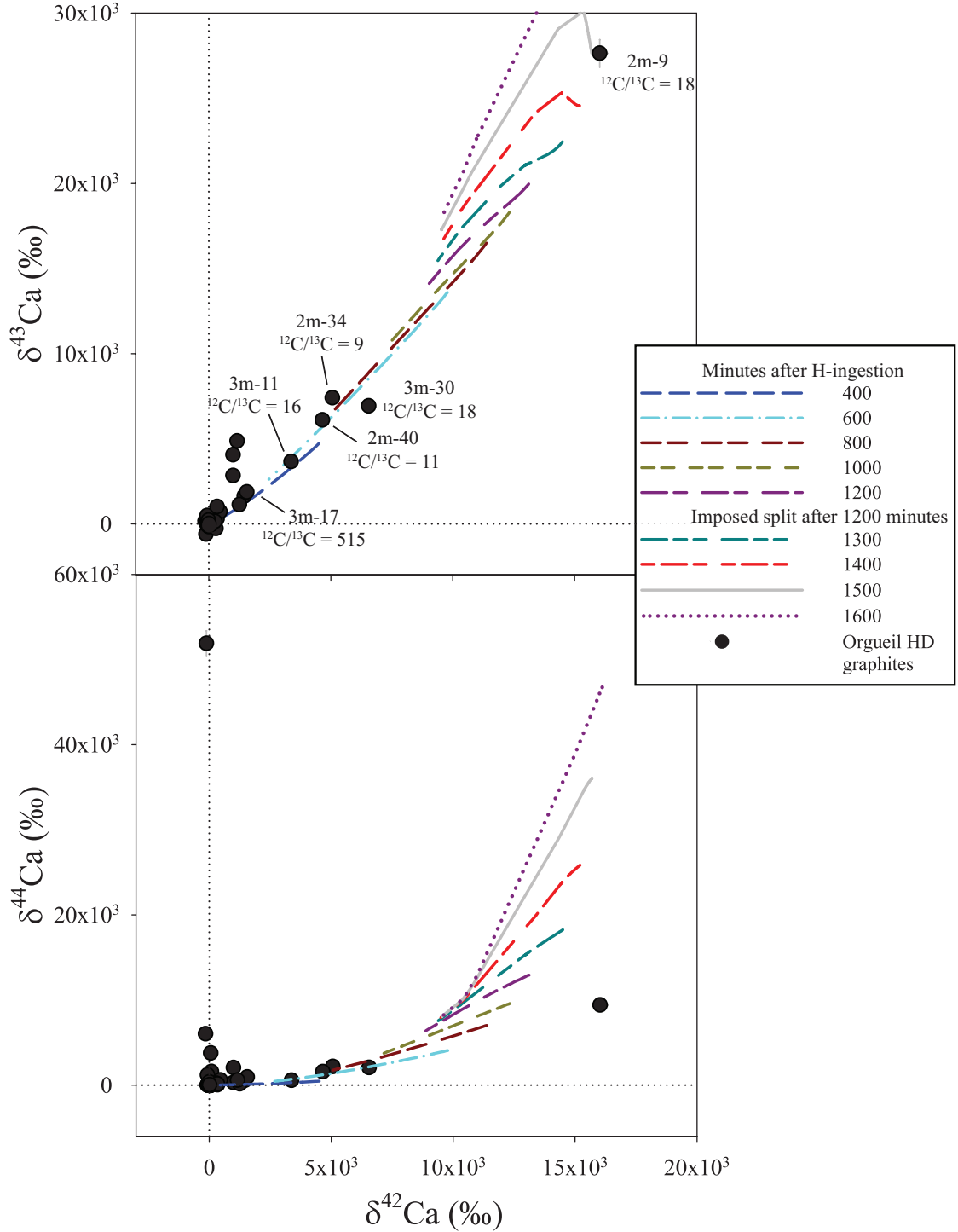}
				\figurenum{3}
        \caption{Three-isotope plots of the $\delta^{42,43,44}$Ca/$^{40}$Ca values measured in HD presolar graphite grains compared with VLTP model predictions by \citet{herwig2011}. In this simulation (RUN106), the He intershell region is split after 1200 minutes. The model calculations agree well with the extreme Ca anomalies observed in the grains. The grains with the largest anomalies are labeled and their $^{12}$C/$^{13}$C ratios are also given.
				Error bars are 1$\sigma$. Dashed lines indicate solar ratios.\label{ca-vltp}}
 \end{figure}

\begin{figure}[h]
       \centering
        \includegraphics[width=27pc]{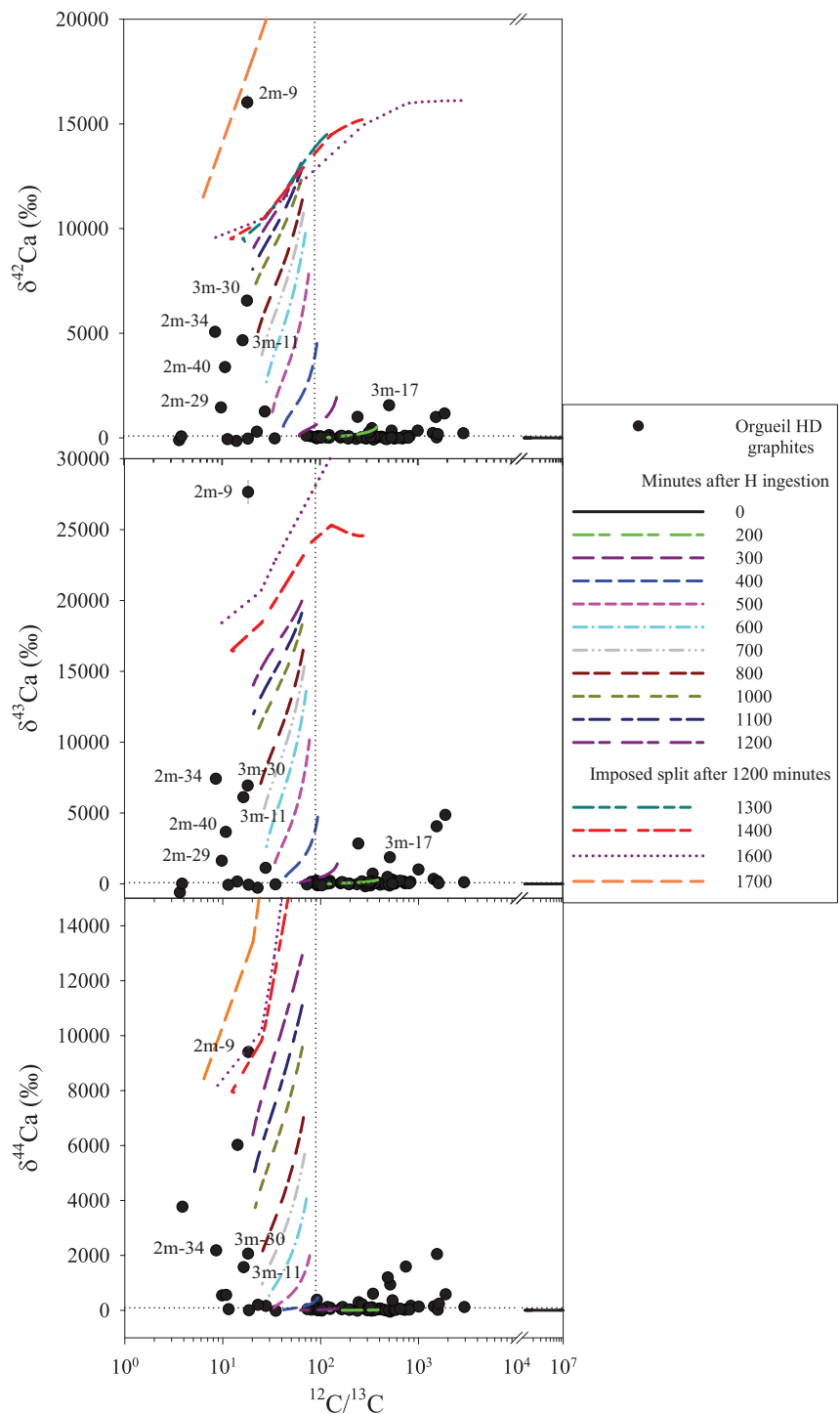}
				\figurenum{4}
        \caption{Ca and C isotopic ratios measured in HD presolar graphite grains compared to VLTP nucleosynthesis model predictions by \citet{herwig2011}. Grains with extreme Ca anomalies are $^{13}$C-enriched while those with moderate anomalies are mostly $^{12}$C-enriched.
				Error bars are 1$\sigma$. Dashed lines indicate solar ratios.
				\label{ca-c-vltp}}
\end{figure}

We found that the Ca isotopic data of the grains agree best with the simulation results for RUN106. The VLTP model calculations are able to explain the large Ca and Ti anomalies observed in grains with low $^{12}$C/$^{13}$C ratios. Figure \ref{ca-vltp} compares Ca grain data with results of RUN106. The grains with the largest Ca anomalies and $^{12}$C/$^{13}$C ratios $<$ 20 (e.g., OR1f2m-9, OR1f2m-34, OR1f2m-40, OR1f3m-30, and OR1f3m-11) fit the model predictions reasonably well. Some grains with $^{12}$C/$^{13}$C ratios $>$ solar and moderate Ca anomalies (e.g., OR1f3m-17) also agree with VLTP calculations. As demonstrated in Figure \ref{cn-vltp}, after the VLTP event, parts of the He intershell can have $^{12}$C/$^{13}$C ratios larger than the solar value. Thus, $^{12}$C-enriched grains with moderate Ca anomalies that are still too high for AGB envelopes and C-rich material from supernova ejecta might also originate in post-AGB stars that have suffered a VLTP. Thus, we see nucleosynthetic signatures from both sides of the split in the presolar graphite grain population. In such a scenario, material from below the split needs to be eventually expelled and mixed with H-burning products from the top of the split into the gas from which the grains condensed. Figure \ref{ca-c-vltp} compares the $^{12}$C/$^{13}$C ratios and $\delta^{42,43,44}$Ca/$^{40}$Ca values of HD graphite grains with those predicted by the models. The grains with extreme $\delta^{42,43,44}$Ca/$^{40}$Ca values are also $^{13}$C-enriched while some grains that have moderate Ca excesses have higher than solar $^{12}$C/$^{13}$C ratios. This supports the hypothesis that $^{13}$C-enriched graphites with extreme Ca and Ti anomalies could have condensed around born-again AGB stars \citep{jadhav2008}. In order for model predictions to agree with Ca and C isotopic data for grains with high $^{12}$C/$^{13}$C ratios (e.g., OR1f3m-17), the Ca excesses have to be diluted without altering the $^{12}$C/$^{13}$C ratios.

\begin{figure}
			\centering
			\begin{subfigure}
        \centering
				\includegraphics[width=17pc]{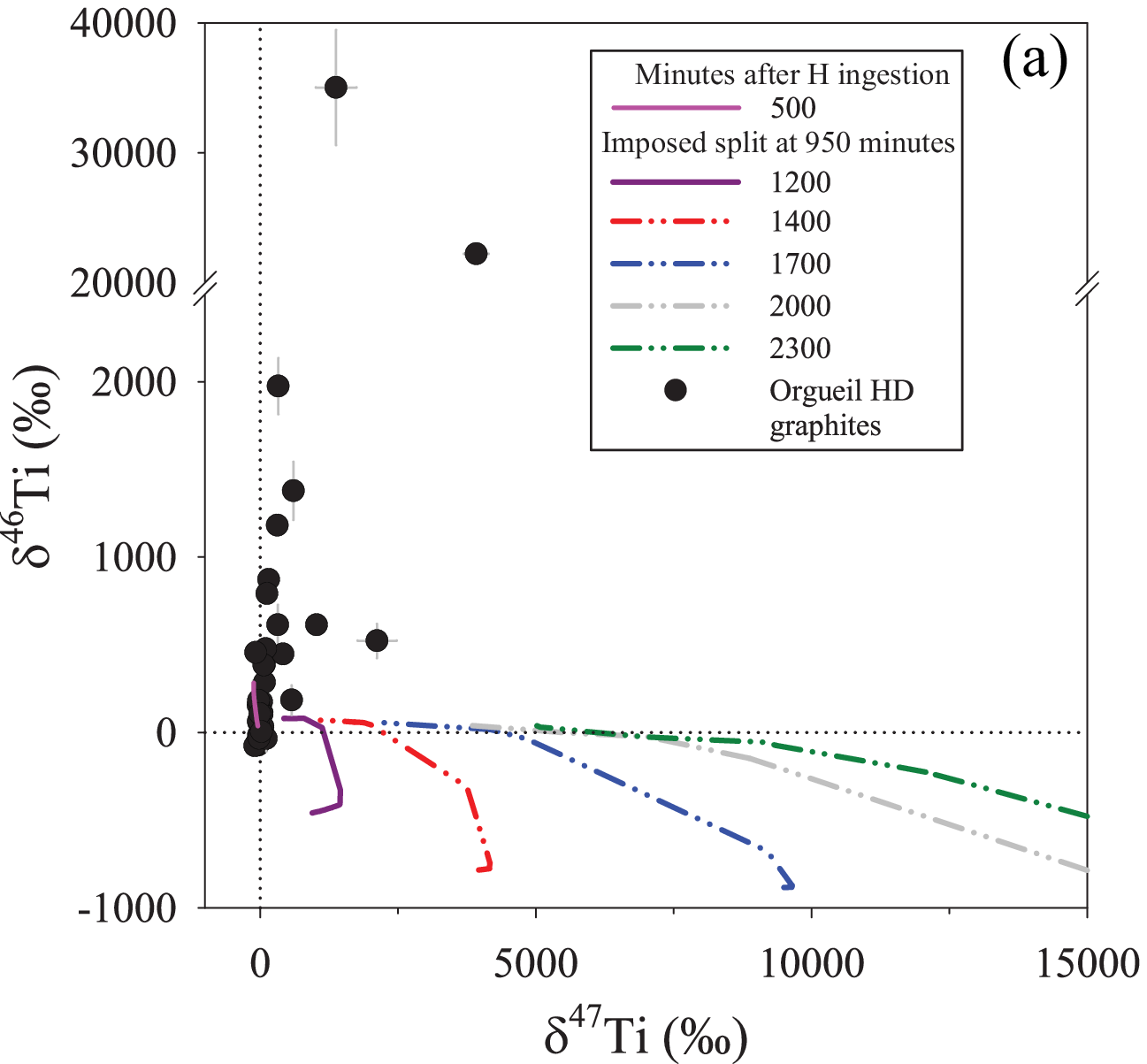}
				\label{pureti-vltp}
      \end{subfigure}
			\begin{subfigure}
        \centering
        \includegraphics[width=19pc]{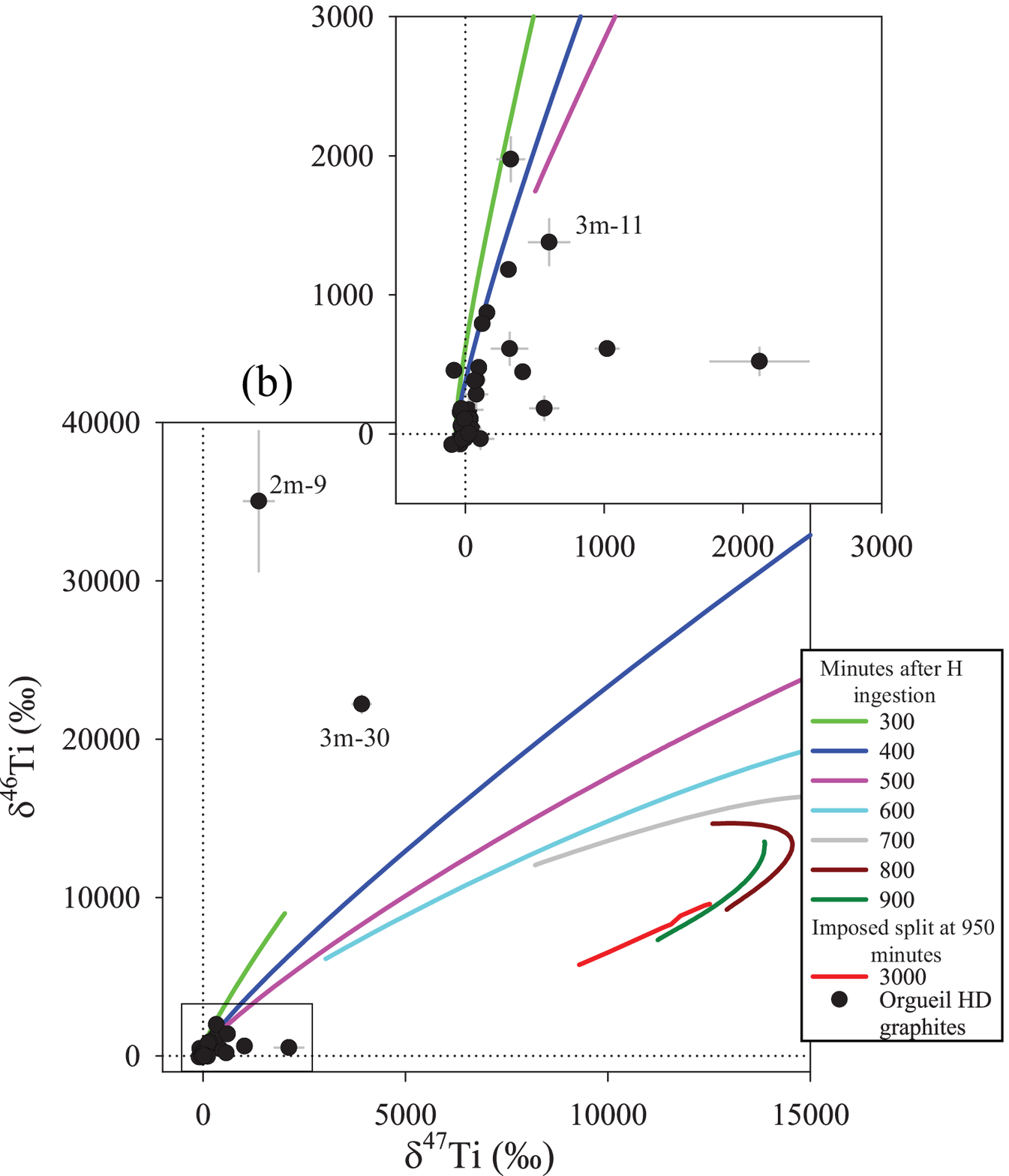}
        \label{caadded-ti-vltp}
			\end{subfigure}
\figurenum{5}
\caption{(a) Three-isotope plot of the $\delta^{46,47}$Ti/$^{48}$Ti values measured in HD presolar graphite grains, compared with VLTP model predictions by \citet{herwig2011}. In this simulation (RUN48), the He intershell region is split at 950 minutes. There is no agreement between grain data and model calculations because $^{46,48}$Ti are destroyed during VLTP nucleosynthesis.
(b) The VLTP calculations (RUN48) shown here now include contributions from $^{46,48}$Ca at masses 46 and 48 that cannot be resolved from $^{46,48}$Ti during the isotopic measurements (in contrast to pure theoretical $\delta^{46,47}$Ti/$^{48}$Ti ratios plotted in Fig. 5a). The model calculations agree with the apparent Ti isotopic ratios observed in some of the grains. The grains with the largest anomalies are labeled to compare with previous Figures 2--4.
				Error bars are 1$\sigma$. Dashed lines indicate solar ratios.}
\end{figure}

Figure \ref{pureti-vltp} compares $\delta^{46,47}$Ti/$^{48}$Ti values measured in grains to model predictions from RUN48. Titanium-46 and $^{48}$Ti are efficiently destroyed during VLTP nucleosynthesis. Thus, the Ti isotopic anomalies measured in graphites do not agree with pure $\delta^{46,47}$Ti/$^{48}$Ti predictions. However, if $^{46,48}$Ca predictions are included with the Ti isotopic predictions (Figure \ref{caadded-ti-vltp}b), then the grain data agree with theoretical values. During SIMS measurements, we are unable to resolve the stable isobars $^{46,48}$Ca from the $^{46,48}$Ti peaks. Thus, we assume that the inferred Ti isotopic ratios can be qualitatively explained by $^{46,48}$Ca excesses in the grains. The high abundances of $^{46,48}$Ca are due to strong neutron capture efficiency during the $i$-process that has been activated by the production of fresh $^{13}$C during the VLTP. $^{46,48}$Ca are not efficiently produced by $s$-process nucleosynthesis during the previous AGB phase. The low $s$-process neutron density does not allow branching on the neutron capture path at the unstable isotopes $^{45}$Ca and $^{47}$Ca, favoring the respective $\beta$-decay channels. Similarly, if the radiogenic contributions from the decay of $^{47,49}$Ca are added to the $^{47,49}$Ti values in the models, then the $^{47,49}$Ti excesses measured in the grains agree better with the VLTP calculations. 

The \citet{herwig2011} VLTP nucleosynthesis models are able to explain both, moderate and extreme Ca and Ti anomalies measured in HD graphite grains that cannot be explained by theoretical predictions for the envelopes of AGB stars and SN ejecta. Grains with the most extreme Ca and Ti ratios are $^{13}$C-enriched and the ones with moderate anomalies can have both high and low $^{12}$C/$^{13}$C ratios.

\section{Discussion}
Low- to intermediate-mass ($0.8-8 M_{\odot}$) AGB stars are the largest contributors of dust to the local interstellar medium \citep{gehrz1989,gail2009,zhukovska2008}. \citet{iben1984} and \citet{renzini1982} estimate that 10\% -- 25\% of stars that leave the AGB track and are on their way to being white dwarfs, undergo a late He flash. Studies of mass loss and dust formation around post-AGB stars that have suffered a VLTP are limited \citep{evans2006,vanhoof2007,chesneau2009} because very few VLTP events have been observed in the high-luminosity phase. Sakurai's object (V4334 Sagittarii) \citep{duerbeck1996} is the most widely studied born-again AGB star; it was discovered while it underwent a VLTP in 1994. Infrared observations of Sakurai's object yield a $^{12}$C/$^{13}$C ratio of $3.5^{+2.0}_{-1.5}$ \citep{worters2009} for the cooling CO ejecta. \citet{eyres1998} and more recently, \citet{chesneau2009} report the presence of thick carbonaceous dust around Sakurai's object. \citet{vanhoof2007} calculate a lower limit for the mass of the total ejecta around Sakurai's object of $6\times 10^{-4} M_{\odot}$ and conclude that this is sufficient to expose the He intershell, confirming the observations of \citet{asplund1999}. The exact mechanism of mixing between the outer layers of the star with the intershell material and whether there is any contribution from the pre-existing planetary nebula are yet to be determined. However, a preliminary comparison of the astronomical observations of Sakurai's object to date, the isotopic data of presolar graphite grains, and VLTP calculations presented in the previous section indicate a scenario in which some presolar graphite grains condensed around similar, old post-AGB stars that suffered a VLTP. The evidence for grains from such stars suggests that we should also be able to find grains from stars that have suffered a late thermal pulse (LTP). A LTP can occur in post-AGB stars while H-burning is still on (e.g., FG Sagittae, Jeffery \& Sch{\"o}nberner 2006; Herwig 2001b). In such stars, material from the He intershell is brought up to the surface during a LTP but there is no ingestion of H into the intershell. Thus, the surface of these stars is $^{12}$C-enriched and exhibits He intershell abundances of Ca and Ti isotopes. In the absence of nucleosynthesis models for such stars, we conjecture that LTP stars could be possible sources of grains that have moderate Ca and Ti anomalies and high $^{12}$C/$^{13}$C ratios.

A comparison of the abundance of graphite grains from post-AGB sources that have suffered a VLTP and the fraction of C dust grains ejected by such objects is beyond the scope of this letter. However, such an investigation will be the next step to determine if VLTP events are the only sources of the C, Ca, and Ti anomalies observed in the grains from this study.  

Most presolar graphite grains with Ti isotopic anomalies contain Ti-rich sub-grains that are the carriers of these anomalies \citep{jadhav2013}. Curiously, the grains with the largest Ca and Ti anomalies presented in this study contain no sub-grains and the Ti anomalies are distributed uniformly throughout the grains. This could be further evidence for rapid condensation in a post-AGB star environment. To further examine our hypothesis, we plan heavy-element (Ba, Mo, Sr) isotopic measurements of these grains which are expected to show strong $i$-process, neutron-burst signatures if the grains originated in a post-AGB source. 

\section{Conclusion}
The comparison of C, Ca, and Ti isotopic data on high-density presolar graphite grains from Orgueil to VLTP nucleosynthesis calculations by \citet{herwig2011} strongly support our previous hypothesis \citep{jadhav2008} that some $^{13}$C-enriched graphite grains with extreme Ca and Ti anomalies could have originated from post-AGB stars that suffered a VLTP. We also found that grains that are $^{12}$C-enriched and contain moderate Ca and Ti anomalies can also condense around stars that suffer a VLTP or LTP. The low C isotopic ratios are a direct signature of H-ingestion, producing $^{13}$C in the upper part of the C-rich intershell. The Ca and Ti anomalies are due, instead, to the activation of the $i$-process at the bottom of the He intershell where neutron densities reach $\sim$ 10$^{15}$ cm$^{-3}$. Our $^{46,48}$Ti isotopic data require that we take into account the contributions of the irresolvable isobars $^{46,48}$Ca to explain the anomalies in the grains. We can verify this assumption in the future by measuring the aforementioned Ca isotopes by Resonant Ionization Mass Spectrometry (RIMS). This technique is capable of ionizing specific elements and suppressing isobaric interferences. Lastly, our study confirms born-again AGB stars as newly identified contributors of dust to the presolar grain inventory and in turn, the solar system. 

The qualitative signature of proton captures coupled with the $i$-process is a robust theoretical prediction of H-ingestion during the VLTP in post-AGB stars. This scenario is also confirmed by independent spectroscopic observations of Sakurai's object. However, we want to highlight the need for multi-dimensional, hydrodynamic simulations to establish stringent, quantitative comparisons between VLTP models and presolar grain data. Such improved hydrodynamic simulations are underway and as an initial step, \citet{woodward2013} have recently shown that quantitative, converged simulations of the H-ingestion process are indeed possible. The new presolar grains diagnostic of the H-combustion event presented here, complements spectroscopic observations. With these multiple constraints, the VLTP evolution phase will provide an excellent laboratory for validating simulations of $i$-process conditions that will be tremendously useful in constructing stellar evolution models for the first generation of stars, where $i$-process conditions are frequently encountered.

\acknowledgments
This work was supported by NASA grants NNX11AG78G (G.R.H.) and NNX11AH14G (EZ). NuGrid acknowledges significant support from NSF grants PHY 02-16783 and PHY 09-22648 (Joint Institute for Nuclear Astrophysics, JINA) and EU MIRG-CT-2006-046520. The continued work on codes and in disseminating data is also made possible through funding from NSERC Discovery grant (FH, Canada) and an Ambizione grant of the SNSF (MP, Switzerland). MP also thanks support from EuroGENESIS. MJ is grateful for Dr. Aurelien Thomen's help with reading and manipulating the VLTP nucleosynthesis files.
\clearpage
\newpage

\clearpage

\clearpage

\clearpage

\clearpage

\clearpage
\end{document}